\documentclass[12pt]{article}

\usepackage{amsmath}
\usepackage[dvips]{graphicx}

\begin{document}

\begin{flushright}
CERN-TH/99-58 \\
MC-TH-99-3\\
March 1999\\
\end{flushright}

\begin{center}
\vspace*{2cm}

{\Large {\bf Extracting the dipole cross-section from photo- and 
electro-production total cross-section data}} \\

\vspace*{1cm}

J.R.~Forshaw$^1$\footnote{On leave of absence from $^2$}, 
G.~Kerley$^2$ and G.~Shaw$^2$

\vspace*{0.5cm}
$^1$Theory Division, CERN,\\
1211 Geneva 23, Switzerland.\\
\vspace*{0.2cm}
$^2$Department of Physics and Astronomy,\\
University of Manchester,\\
Manchester. M13 9PL. England.

\end{center}
\vspace*{4cm}

\begin{abstract}
We report on a successful attempt to extract the cross-section for
the high-energy scattering of colour dipoles of fixed transverse size off 
protons using electroproduction and photoproduction total cross-section data,
subject to the constraint provided by the ratio of the overall photon
dissociation cross-section to the total cross-section.
\end{abstract}

\mbox{13.60.Hb, 25.20.Lj, 25.30.Rw, 13.38.-t, 24.85.+p, 12.40.Nn}

\newpage

\section{Introduction}

Photon-hadron interactions at high energies can be described in the rest frame
 of the hadron using a picture in which the incoming photon undergoes a 
fluctuation into virtual partonic or hadronic states, which subsequently 
interact strongly with the hadron. A number of singly-dissociative diffractive
 processes, namely elastic Compton scattering,  photon dissociation, exclusive
 vector meson production and deeply virtual Compton scattering, can be 
formulated in terms of a quantity, the \emph{colour dipole cross-section}, 
which is universal for a given hadron target~\cite{dcs_nik, pom_DD_nik, 
dip_muel_1, dip_muel_2}.
We report on a successful attempt to extract the cross-section for scattering
colour dipoles of fixed transverse size off protons using both
electroproduction and photoproduction $\gamma p$ total cross-section data, 
subject to the constraint provided by the ratio of the overall photon
dissociation cross-section to the total cross-section.

We begin by briefly summarising the colour dipole model of diffraction;
we then describe the assumed forms of the dipole 
cross-section and the photon wave functions before discussing the data fits 
and the resulting values of the dipole cross-section.

\section{Colour dipoles and diffraction}

\subsection{The $\gamma p$ total cross-section}

Our first task is to introduce the colour dipole cross-section and relate it 
to the $\gamma p$ total cross-section. Here we follow closely the treatment of
 ~\cite{forshaw1, forshaw3}. 
Of particular utility in the study of diffractive scattering is a 
decomposition of the strongly interacting fluctuations of the   photon into a 
superposition of Fock states in the quark-gluon basis:
\begin{equation}
   |\gamma\rangle = \sum |q\bar{q}\rangle + |q\bar{q}g\rangle + \mbox{ higher
    Fock states}  \; \; .
\end{equation}

\begin{figure}[htbp]
  \begin{center}
    \includegraphics[width=14cm,height=10cm]{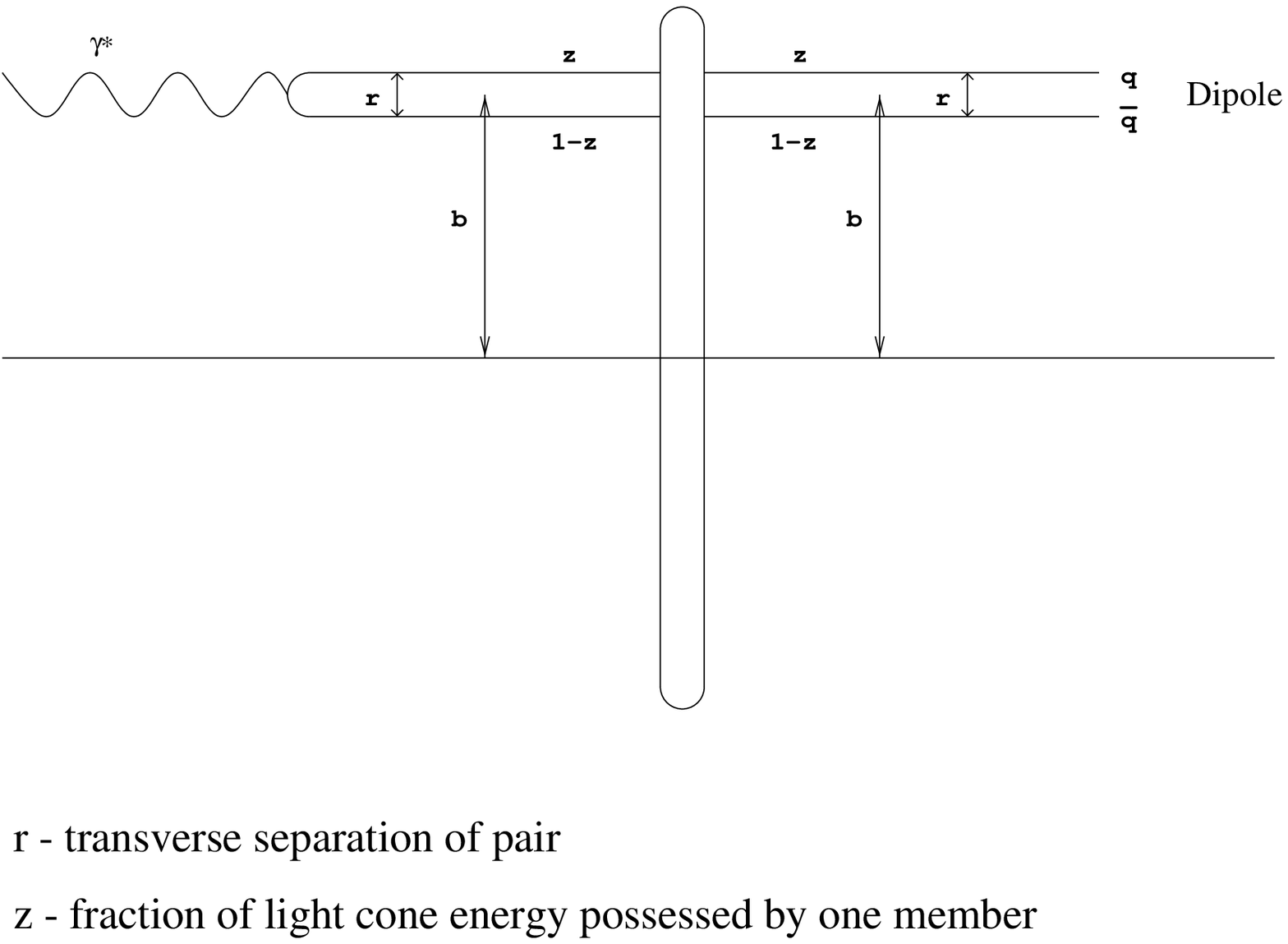} 
    \caption{The diffractive process from a mixed position-momentum
 viewpoint. Transverse components are spatial; non-transverse components are 
light cone momenta.}
    \label{fig:diff_op}
  \end{center}
\end{figure}

We define $r$ as the transverse separation averaged over all orientations of
 the quark-antiquark pair and $z$ as the fraction of the light-cone energy of
 the photon carried by one of the pair (Fig.~\ref{fig:diff_op}). 
Quark-antiquark states with definite values of $z$ and $r$ preserve these 
values during the diffractive process; or, to put it another way, they are 
eigenstates of the scattering matrix $\hat{T}$ when it is restricted to 
diffractive processes.  This we shall call the diffraction operator. The 
quark-antiquark eigenstates are called \emph{colour dipoles}.
Expanding the virtual photon in these states gives
\begin{equation}
  \label{eq:photon_wf}
  |\gamma\rangle = \int \mbox{d}z \mbox{d}^{2}r \ \psi (z,r) |z,r\rangle + 
\mbox{ higher Fock states,}
\end{equation}
where $\psi (z,r)$ is the \emph{light cone wave function} of the photon.

The diffractive process modulates the light cone wave function by the 
eigenvalue $\tau$ of the diffraction operator:
\begin{equation}
  \label{eq:eigen_val}
  \hat{T} |z,r\rangle = i \tau(b,s; z,r) |z,r\rangle.
\end{equation}
Here $b$ is the \emph{impact parameter} of the dipole with respect to the 
proton centre, being a weighted sum of the individual impact parameters $b_{1}
, b_{2}$ of its
constituents:~\cite{vm_dosch_1} 
\begin{eqnarray}
  \label{eq:b_def}
  b & = & |\mathbf{b}|  \nonumber \\
  \mathbf{b} & = & z \mathbf{b}_{1} + (1-z) \mathbf{b}_{2}.
\end{eqnarray}
The factor $i$ in (\ref{eq:eigen_val}) is inserted for convenience; it ensures
 that $\tau$ is predominantly real (since diffractive amplitudes are 
predominantly imaginary).

Consider the $\gamma^* p$ total cross-section in deep inelastic scattering. 
We first express the elastic scattering amplitude for $\gamma^* p \to 
\gamma^* p$ in terms of the Mandelstam variables 
$s = W^{2}$ and $t$, which can be done by a Fourier transform with respect to 
the momentum conjugate to $b$, namely the perpendicular part of the proton 
momentum transfer: $q' = p - p'$. So we arrive at
\begin{equation}
  \label{eq:scatt_amp}
  A^{el}(s, t) = \int \mbox{d}^{2}b \ e^{i\mathbf{q'_{\perp}\cdot b}}\langle
\gamma|\hat{T}|\gamma\rangle.
\end{equation}
Use of the optical theorem leads to
\begin{equation}
  \label{eq:sigma_tot}
   \sigma^{\gamma^{*}p}_{T,L} = \int \mbox{d}z \mbox{d}^{2}r \ |\psi_{\gamma}^
{T,L}(z,r)|^{2}\int\frac{\mbox{d}^{2} b \ \tau(s,b;z,r)}{s}. 
\end{equation}
The second integral expression defines the colour dipole cross-section:
\begin{equation}
  \label{eq:cross_sec_def}
  \int\frac{ \mbox{d}^{2} b \ \tau(s,b;z,r)}{s} \equiv \sigma(s,r,z).
\end{equation}
This is the total cross-section for scattering dipoles of a specified 
configuration $(z,r)$ off a proton.

\subsection{Other processes}

There are other interesting processes which involve the dipole cross-section: 
vector meson production and photon dissociation.
The formulation of the first is straightforward. The differential 
cross-section is given by:
\begin{equation}
  \label{eq:vector_mes}
  \left. \frac{d \sigma^{V}_{T,L}}{dt}\right|_{t=0} = \frac{1}{16 \pi}\left 
[\int dz d^{2}r \ \psi_{V}^{*}(z,r) \psi_{\gamma}^{T,L}(z,r) \sigma(s,r,z) 
\right ]^{2}
\end{equation}

For photon dissociation, we can express the final state as an incoherent sum 
of the diffractive eigenstates (dipole states):~\cite{forshaw3}
\begin{equation}
  \label{eq:phot_diss_fin_st}
  \left.\frac{d \sigma^{D}_{T,L}}{dt}\right|_{t=0} = \frac{1}{16 \pi s^{2}}
\sum_{k}|\langle\gamma^{T,L}|\hat{T}|z,r\rangle|^{2}
\end{equation}
and hence 
\begin{equation}
  \label{eq:phot_diss}
   \left.\frac{d \sigma^{D}_{T,L}}{dt}\right|_{t=0} = \frac{1}{16 \pi} \int dz
 d^{2}r |\psi_{\gamma}^{T,L}(z,r)|^{2}\sigma^{2}(s,r,z).
\end{equation}
Note that only that subset of the diffractive dissociation final state which 
is composed exclusively of a quark--anti-quark pair has been included in this 
expression.

The dipole cross-section thus constitutes a link between three distinct 
physical processes.  If the dipole cross-section is known, then vector meson 
wave functions  predicted from models can be inserted in (\ref{eq:vector_mes})
 and the models tested by comparison of the result with experiment. 
Alternatively, the vector meson wave function itself can be extracted.

\section{Parametric forms}

In what follows, our aim is to extract the dipole cross-section  from total
cross-section data for virtual photoabsorption by protons (structure function
and real photoabsorption data); and to use the result to predict the
contribution to the photon dissociation rate from dipole scattering.  
In order to do this, it is
necessary to assume parametric forms for  the dipole cross-section which embody
reasonable theoretical requirements, but are otherwise flexible. Here we
describe the form used in our fits, together with our assumptions for the
photon  wavefunction.

\subsection{The dipole cross-section}

The dipole cross-section is in general a function of $z$, $s = W^2$ and $r$.
However 
a non-perturbative calculation reveals little $z$ dependence~\cite{vm_dosch_1}
and we shall neglect it completely in what follows.

Following other authors~\cite{Moseley, double_pom} we assume the existence of 
two distinct terms
 which carry a Regge type $s$ dependence: the hard term, which is assumed to
dominate at small $r$ but vanish in the limit of large $r$; and the soft term,
 with an $s$ exponent close to zero, which is assumed to dominate at large $r$
 and saturate. Specifically, we assume 
\[
  \sigma(s,r)  =  \sigma_{soft}(s,r) + \sigma_{hard}(s,r) \ 
\]
where the $r$-dependences are given by 
\begin{eqnarray}
  \label{eq:new_par}
  \sigma_{soft}(s,r) & = & a_{0}^{S}\left(1 - \frac{1}{1 + (a_{1}^{S}r + 
a_{2}^{S}r^{2})^{2}}\right)(r^{2}s)^{\lambda_{S}} \nonumber \\
   \sigma_{hard}(s,r) & = & (a_{1}^{H}r + a_{2}^{H}r^{2} + a_{3}^{H}r^{3})^{2}
  \exp(-\nu_{H}^{2}r) (r^{2}s)^{\lambda_{H}} \; .
\end{eqnarray}
In these formulae the energy variable $s$ has an associated  $r^{2}$ factor,
which  yields an 
implicit $Q^{2}$ dependence and approximate scaling on integrating over the 
photon
wavefunction.
 Apart from this, both terms possess a limiting $r^{2}$ dependence at small $r$
in accordance with colour transparency\footnote{Colour transparency arguments 
usually assume
a parameterisation in  $x$,  which contains an 
implicit $Q^{2}$ dependence, rather than $s$. 
It is unnatural to have a $Q^{2}$ dependence in 
the dipole cross-section itself~\cite{low_q_par_ros} and  
we prefer to introduce it as the transform of an additional $r$ dependence via 
the photon wave function.} 
 arguments~\cite{col_tra_nik}.  The
squared polynomials in $r$ provide a fine tuning of the $r$ dependence that is
 strictly non-negative.

\subsection{The photon wavefunction}

In the first instance, we used the tree level QED form of the photon light cone
wave function:
\begin{eqnarray}
  \label{eq:psi^2}
  |\psi_{L}(z,r)|^{2} & =  & \frac{6}{\pi^{2}}\alpha\sum_{q=1}^{n_{f}}e_{q}^
{2}Q^{2}z^{2}(1-z)^{2} K_{0}^{2}(\epsilon r) \\
  |\psi_{T}(z,r)|^{2} & = & \frac{3}{2 \pi^{2}}\alpha\sum_{q=1}^{n_{f}}e_{q}^
{2} \left\{[z^{2} + (1-z)^{2}] \epsilon^{2} K_{1}^{2}(\epsilon r) + m_{f}^{2} 
K_{0}^{2}(\epsilon r) \right\}
\end{eqnarray}
where
\[
 \epsilon^{2} = z(1-z)Q^{2} + m_{f}^{2}\; ,
\]  
 $K_{0}$ and $K_{1}$ are modified Bessel functions and the sum is over quark
flavours.  The quark masses can be 
neglected at large $Q^{2}$, but are important at low $Q^2$. Here we assume
three light quark flavours with a
generic value $m_f^2 = 0.08$ GeV$^2$. This corresponds roughly to a constituent
 quark mass and enables  good fits to real as well as virtual photon data
to be obtained. The use of a constituent as opposed to a current quark mass 
can be regarded as a partial reflection of confinement. Subsequently we 
found it necessary to incorporate other confinement effects in the 
wavefunction,
as described below.

Finally, the  absence of a $z$ dependence in the dipole cross-section allows us
to explicitly integrate over it in (\ref{eq:sigma_tot}) to give
\begin{eqnarray}
\label{eq:z_int}
   \sigma^{\gamma^{*}p}_{tot} & = & \int \mbox{d}z \ \mbox{d}^{2}r \
( |\psi_{T}(z,r)|^{2} + |\psi_{L}(z,r)|^{2} ) \sigma(s,r) \nonumber \\
   & = &  \frac{6}{\pi^{2}}\alpha\sum_{q=1}^{n_{f}}e_{q}^{2} \int  
\mbox{d}^{2}r \frac{G(r)}{r^{2}} \sigma(s,r), 
\end{eqnarray}
for $ \sigma = \sigma_T +  \sigma_L $, where 
\begin{eqnarray}
  \label{eq:G_def}
  \lefteqn{G(r) =}   \nonumber \\ & & \hspace*{-0.5cm}
\int_{0}^{1} \mbox{d}z r^{2} \left\{[Q^{2}z^{2}(1-z)^{2}  + \frac{m_{f}^{2}}
{4}]K_{0}^{2}(\epsilon r)  + \frac{[z^{2} + (1-z)^{2}] \epsilon^{2} K_{1}^{2}
(\epsilon r)}{4} \right\}.
\end{eqnarray}

\section{Extracting the dipole cross-section}

\subsection{The data set}

The $F_{2}$ data set consisted of HERA 1994 and 1995 data from the
H1~\cite{dat:H1_94, dat:H1_LOQ_95} and ZEUS~\cite{dat:ZEUS_94, dat:ZEUS_2_94,
dat:ZEUS_BPC_95, dat:ZEUS_SVX_95} experiments, together with the fixed target
E665 values~\cite{dat:E665}.  This was combined  with the very precise
intermediate  energy photoproduction data~\cite{dat:cald} plus the two high 
energy 
points from H1~\cite{dat:H1_phot} and ZEUS~\cite{dat:ZEUS_phot}
respectively. The 
following cuts were imposed:
\begin{itemize}
\item  A cut in $s$ ($s \ge 100 \mbox{ GeV}^{2}$) to ensure the data was 
sufficiently high energy.
\item A cut in $x$  ($0 \le x \le 0.01$) to ensure the data was diffractive. 
\item  A cut in $Q^2$  ($Q^2 \le 60 \mbox{ GeV}^{2}$). 
\end{itemize}
Altogether there were 345 $F_{2}$ and  20 photoabsorption data points, compared
to 10 adjustable parameters in our final fits, described below. 

In fitting these data, the purely diffractive contribution described above  
was supplemented by a
small non-diffractive component arising from the leading meson exchange 
trajectories. This was assumed to be given by the empirical
Donnachie--Landshoff form:~\cite{DL_fit}
\begin{eqnarray}
  \label{eq:DL_regge}
  F_{2}^{R} & = & 0.098\ x^{0.4525} \left( \frac{Q^{2}}{Q^{2} + 0.0111} 
\right)^{0.5475} \mbox{  (electroproduction)}\nonumber \\
  \sigma_{\gamma p}^{R} & = & 0.3318 \ s^{-0.4525} \mbox{ GeV}^{-2}   \mbox{  
(photoproduction)}.
\end{eqnarray}
Its contribution was always less than 15\% for photoproduction and typically 
3\% or less for electroproduction.

\subsection{The fits}

The inversion of integral equations of the type in (\ref{eq:z_int}) presents 
notorious problems of non-uniqueness and instability (sensitivity to small 
alterations in input data) of the resulting 
function ~\cite{inv_ast, gam_burst}.  
Certain features of our fits, however, mitigate these effects.  
Our parameterization ensures that the dipole cross-section is strictly 
positive, as is essential from its physical interpretation, and this 
already goes a long way to ensure that the output is robust towards data 
fluctuations. Further, we have imposed many additional constraints from 
physical considerations which limit the degree of arbitrariness in the final 
fit.
Nevertheless, we were unable to achieve a positive definite error matrix for 
our fits so that the errors quoted below are approximate.

\subsubsection{Fits with a QED wavefunction}
Using the QED photon wave functions of (\ref{eq:psi^2}) led to successful fits
to the  $\gamma p $ total cross-section data using the above and other
similar parameterizations. However, although the $\chi^{2}$ values typically
ranged from 0.9 to 1.2  
per degree of freedom, this was achieved 
at the price of  unphysically large dipole cross-sections for dipole sizes 
greater than or of order 1 fm. For example, these were found 
to be of order 100~mb at $\surd s = 100 \mbox{ GeV}$, compared with a
 $\rho^0 N$ cross-section of about  25 mb ~\cite{rhoN}.

 Such a large dipole cross-section has the effect of predicting much too high 
a rate for diffractive processes, which are more sensitive to large dipoles 
since the dipole cross-section is squared in (\ref{eq:phot_diss}).  
This rate can be calculated from our parameterisation 
using~(\ref{eq:phot_diss}) and integrating over $t$ using the relation
\begin{equation}
  \label{eq:t_dep}
   \left.\frac{d \sigma^{D}}{dt}\right| =  \left.\frac{d
   \sigma^{D}}{dt}\right|_{t=0} \exp(-b |t|) \; ,
\end{equation}
where for  the slope parameter $b$ we used the value 
7.2 GeV$^{-2}$~\cite{F2D4}.
This  leads to  predicted diffractive cross-sections which are typically 
over 45\% of the total for real photons.
In contrast, the experimental values are 
 $22.2 \pm 3.2 \%$ $(s = 3.5 \times 10^{4} \mbox
{ GeV}^{2})$~\cite{adloff.dissoc} according to the H1 Collaboration and $13.3
\pm 3.6 \%$ $(s = 4 \times 10^{4} \mbox{ GeV}^{2})$~\cite{breitweg.dissoc}
according  to the ZEUS Collaboration.

Thus, in spite of the apparent success of the fits, there are clearly serious 
shortcomings in the above approach. This is confirmed by both the flexibility 
of the functional forms chosen, and
the fact that we could not fit the data at all when we impose reasonable
limits on the cross-sections for large dipoles. The obvious suspect is the
assumed form of the photon wave function at large transverse size,
where confinement effects are surely significant.
 
\subsubsection{Modifying the photon wave function}

We  adopt a pragmatic, \emph{a posteriori} approach to this problem by 
modifying the photon wave function so that the soft contribution to the 
dipole cross-section is brought into line with the above experimental 
constraints, while the  hard contribution is unaltered.

As mentioned above, the high value for the diffractive to total cross-section 
ratio is indicative of inflated values of the dipole cross-section at 
large $r$.  If  the photon wave function were larger at those large $r$ 
values for which the integrand of (\ref{eq:sigma_tot}) is still appreciable, 
then the value of the diffractive cross-section would be smaller.
Consequently, we multiply $G(r)$ by a shifted Gaussian: 
\begin{equation}
  \label{eq:peak}
  f(r) = \frac{1 + B \exp(- c^{2} (r - R)^{2})}{1 + B \exp(- c^{2} R^{2})}.
\end{equation}
This form enables the width and height of the enhancement to be controlled 
independently while keeping a factor of close to 
unity at small $r$. 

\begin{figure}[htbp]
  \begin{center}
   \includegraphics[width=16cm,height=16cm]{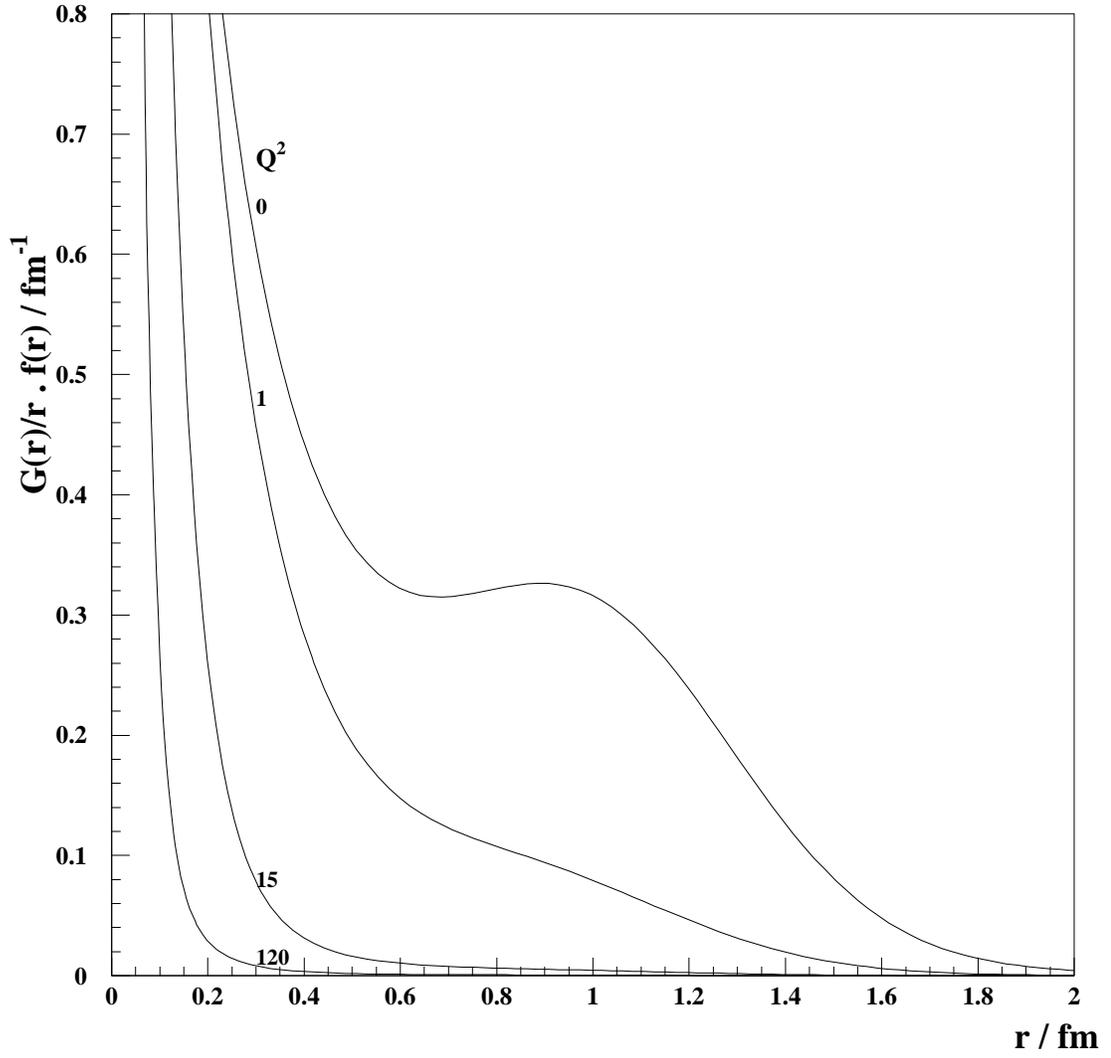}    
    \caption{The weight function $f(r) G(r) / r$ for 
different $Q^{2}$ (fit I). 
The peak at low $Q^{2}$ represents the modification to the photon 
wave function.}
    \label{fig:40.657.gr}
  \end{center}
\end{figure}

The resulting behaviour of $G(r)$ is shown in Fig.~\ref{fig:40.657.gr}
for the parameter values of our final fit I, described below. The 
behaviour at both small and large $r$-values is very similar to 
that suggested by a  successful ``off-diagonal''  generalised vector 
dominance model~\cite{GVD}, where the 
probability distribution of scattering eigenstates  
exhibits peaks for cross-sections of hadronic size on an otherwise monotonic 
decrease with $\sigma$~\cite{x_fluct}.

\subsubsection{Fits with the modified wave function}

On refitting the data, we were able to adjust the ratio of diffractive to 
total cross-section for photoproduction to any reasonable value by adjusting
the value of the saturation parameter $a_{0}^{S}$. In addition, we 
fixed the value of the exponent $\lambda_{S}$ to ensure reasonable
agreement with the high energy real photoabsorption data points, which are
of low statistical significance, as described below.
Two fits I and II are
reported here and summarised in Tables 1 and 2. They differ in that they give 
diffractive ratios at $\sqrt{s}$ of 180 GeV of 14\% and 23\% to agree to
 within 2\%
with ZEUS and H1 photoproduction values respectively. 
Since the fits are similar, we shall concentrate on Fit I, commenting 
briefly on the comparison with fit II where appropriate.

\begin{figure}[htbp]
  \begin{center}
   \rotatebox{90}{\includegraphics[width=12cm,height=16cm]{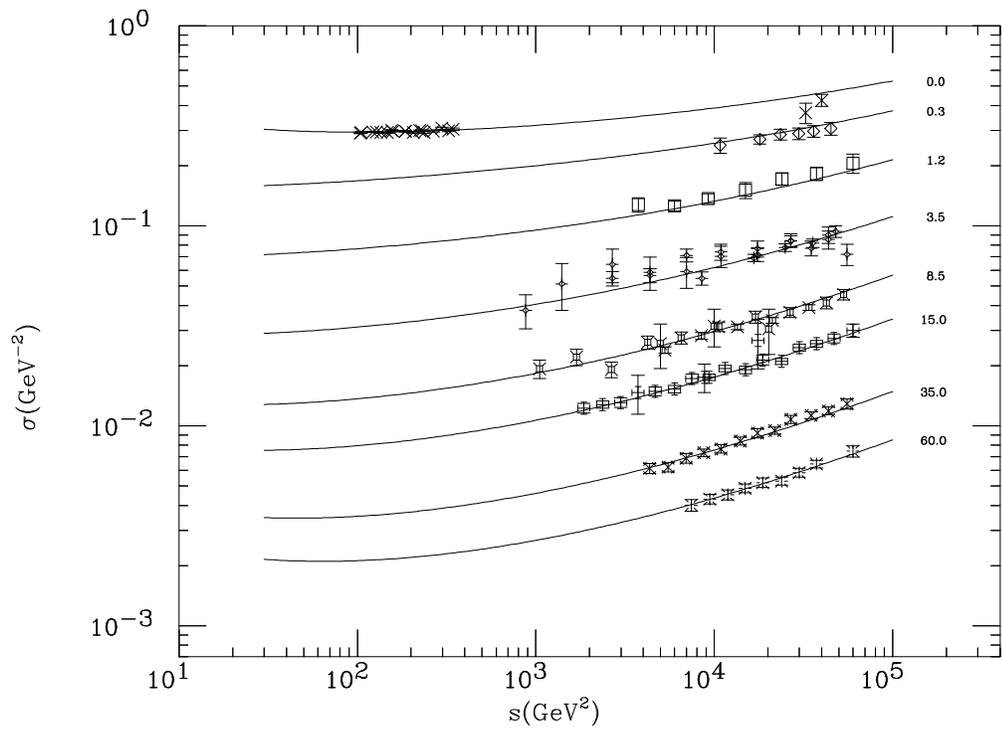}}
    \caption{ Representative sample of fitted data points for 
the total cross-section $\sigma_{\gamma p}^{tot}$ compared with curves 
calculated from the parameterised dipole cross-section for different 
 $Q^{2}$ values (fit I).}
    \label{fig:40.657.plot}
  \end{center}
\end{figure}

The quality of the fit is illustrated in Fig.~\ref{fig:40.657.plot}. 
The fit has a good $\chi^{2}$ but not so low as to indicate overfitting 
and the contribution from the very precise 
intermediate photoabsorption data is reasonably small.  At high energies,
the  photoabsorption total cross-section  lies somewhat
above the ZEUS point especially, even though the soft 
term $s$ exponent $\lambda_{S}$ was given a slightly low value 0.06 compared 
to the canonical Donnachie-Landshoff~\cite{DL_fit} value of  0.08
 to improve the agreement.  However, these data values are low in comparison 
with  a generalised vector dominance based extrapolation from 
low $Q^{2}$ ZEUS data~\cite{Xsec_born}.
As regards the hard term, its $s$ exponent $\lambda_{H}$ is consistent with 
the `hard pomeron' intercept of 1.418 obtained by Donnachie and 
Landshoff~\cite{double_pom}. 

\begin{figure}[htbp]
  \begin{center}
   \includegraphics[width=16cm,height=16cm]{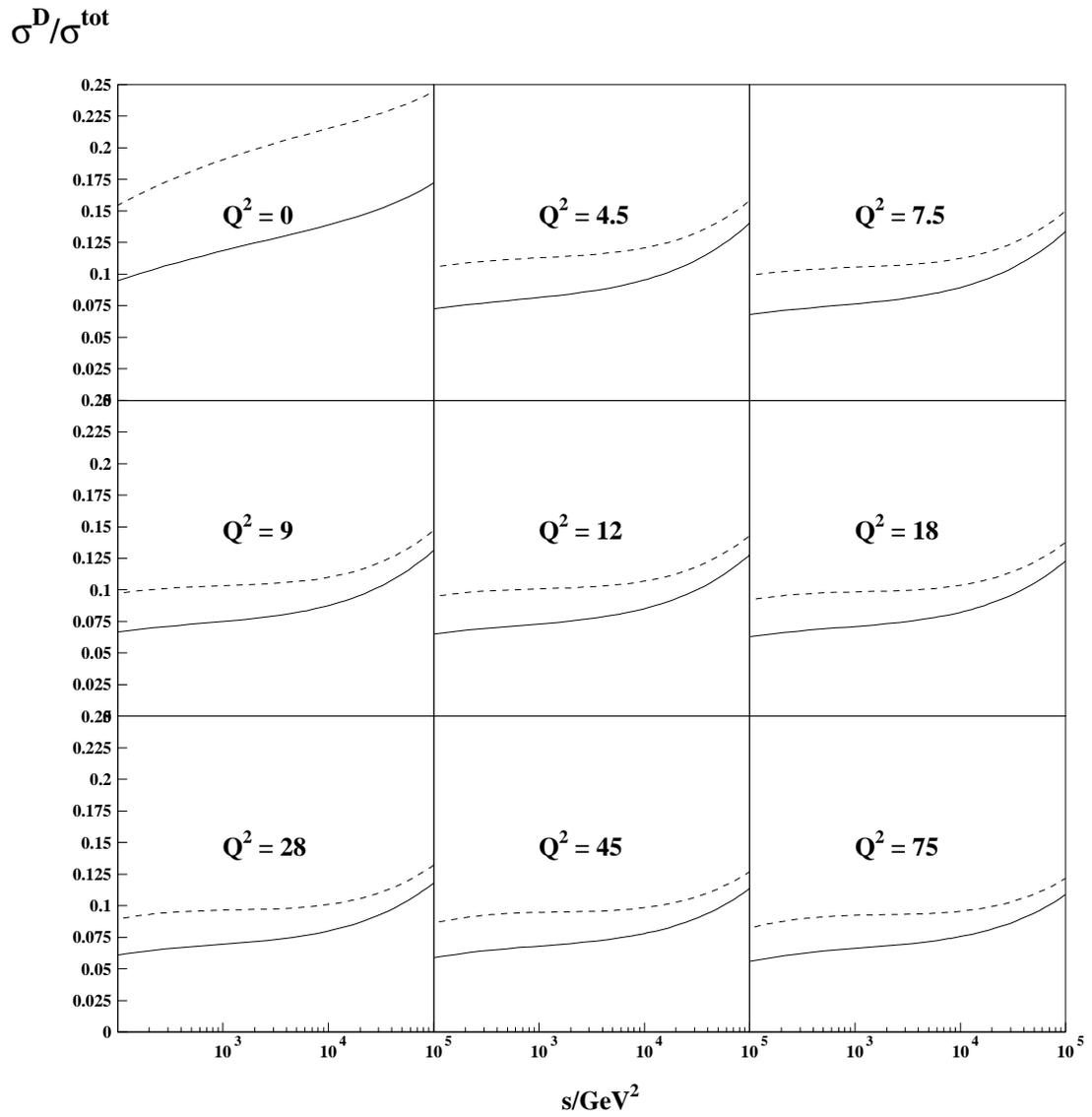}     
    \caption{ Ratio of the overall singly dissociative 
diffractive cross-section to the total cross-section for 
fit I (solid line) and fit II (dotted line).}
    \label{fig:40.1054.825.dtot}
  \end{center}
\end{figure}

The predictions for the ratio of diffractive to total cross-section are shown 
in  Fig.~\ref{fig:40.1054.825.dtot}. As can be seen, there is  little 
variation 
with $Q^{2}$, which accords with the $Q^{2}$ independence 
of $F_{2}^{D(2)}(\beta,Q^{2})$. The weak $s$ dependence is also in 
line with experiment~\cite{high_doy}.

\subsubsection{The dipole cross-section}

\begin{figure}[htbp]
  \begin{center}
   \includegraphics[width=16cm,height=16cm]{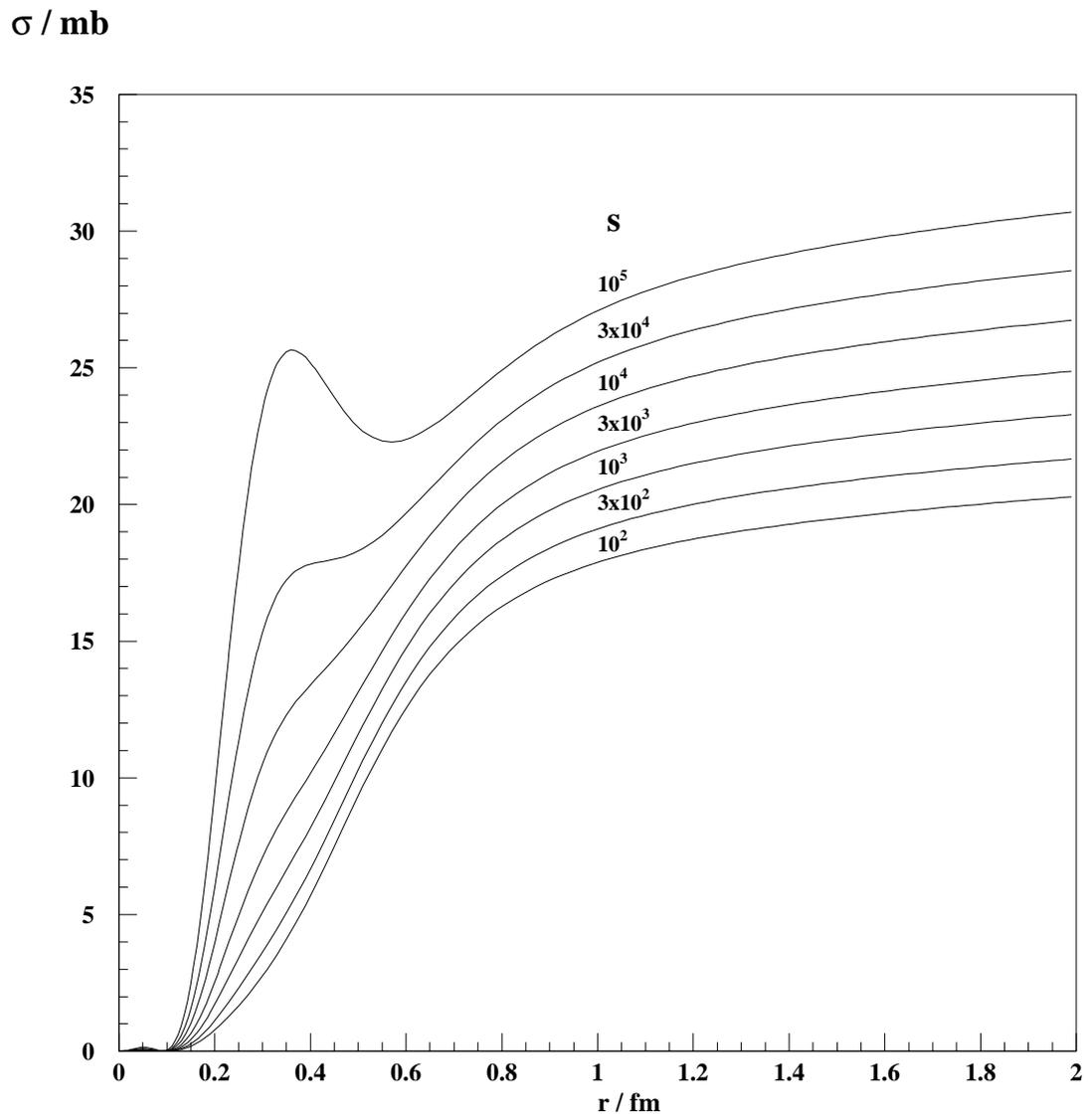}    
    \caption{The dipole cross-section at different energies (fit I).}
    \label{fig:40.657.dcs}
  \end{center}
\end{figure}

\begin{figure}[htbp]
  \begin{center}
   \includegraphics[width=16cm,height=16cm]{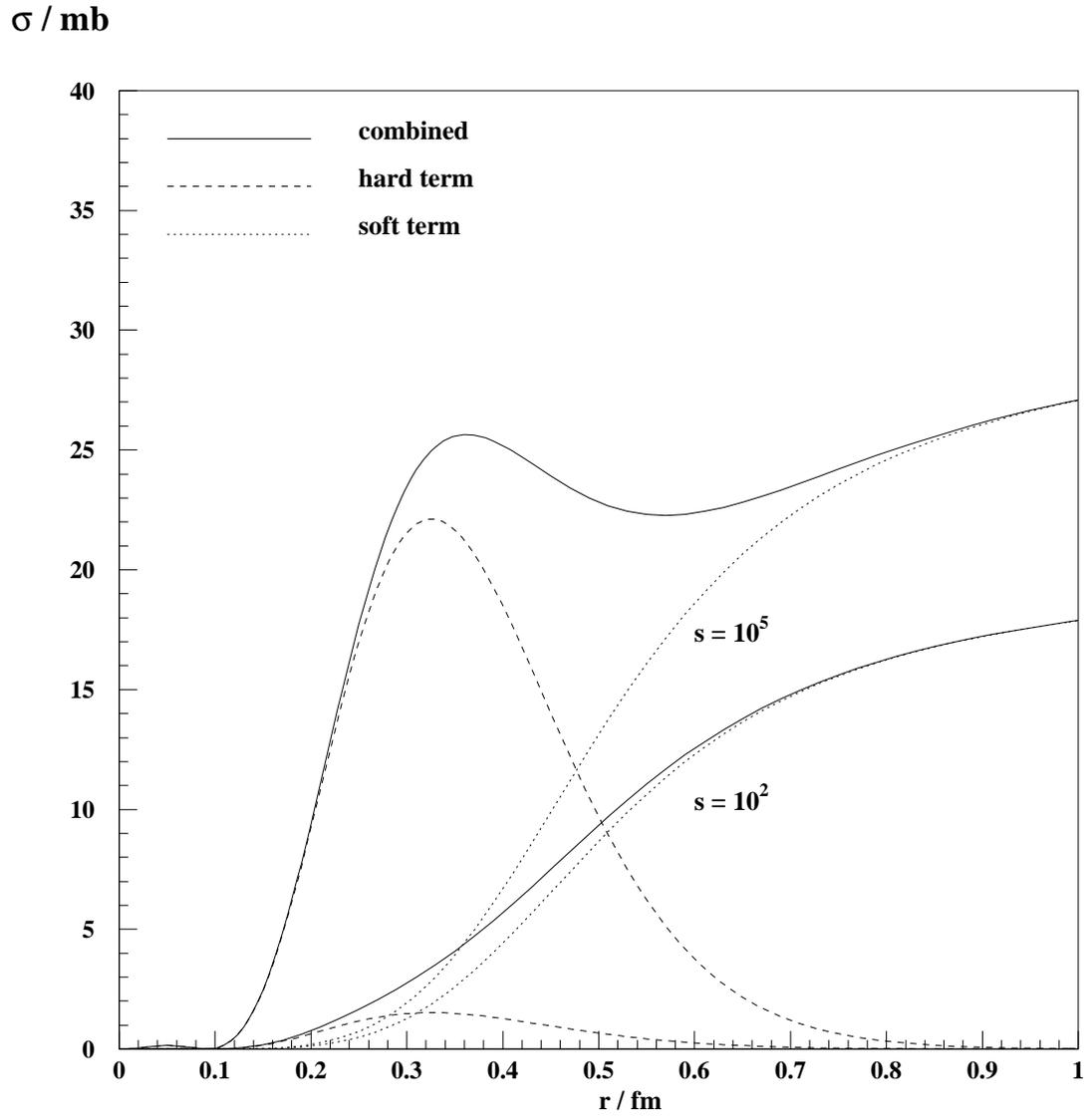}    
    \caption{Hard and soft contributions to the dipole cross-section (fit I).}
    \label{fig:40.657.hs}
  \end{center}
\end{figure}

\begin{figure}[htbp]
  \begin{center}
   \includegraphics[width=16cm,height=16cm]{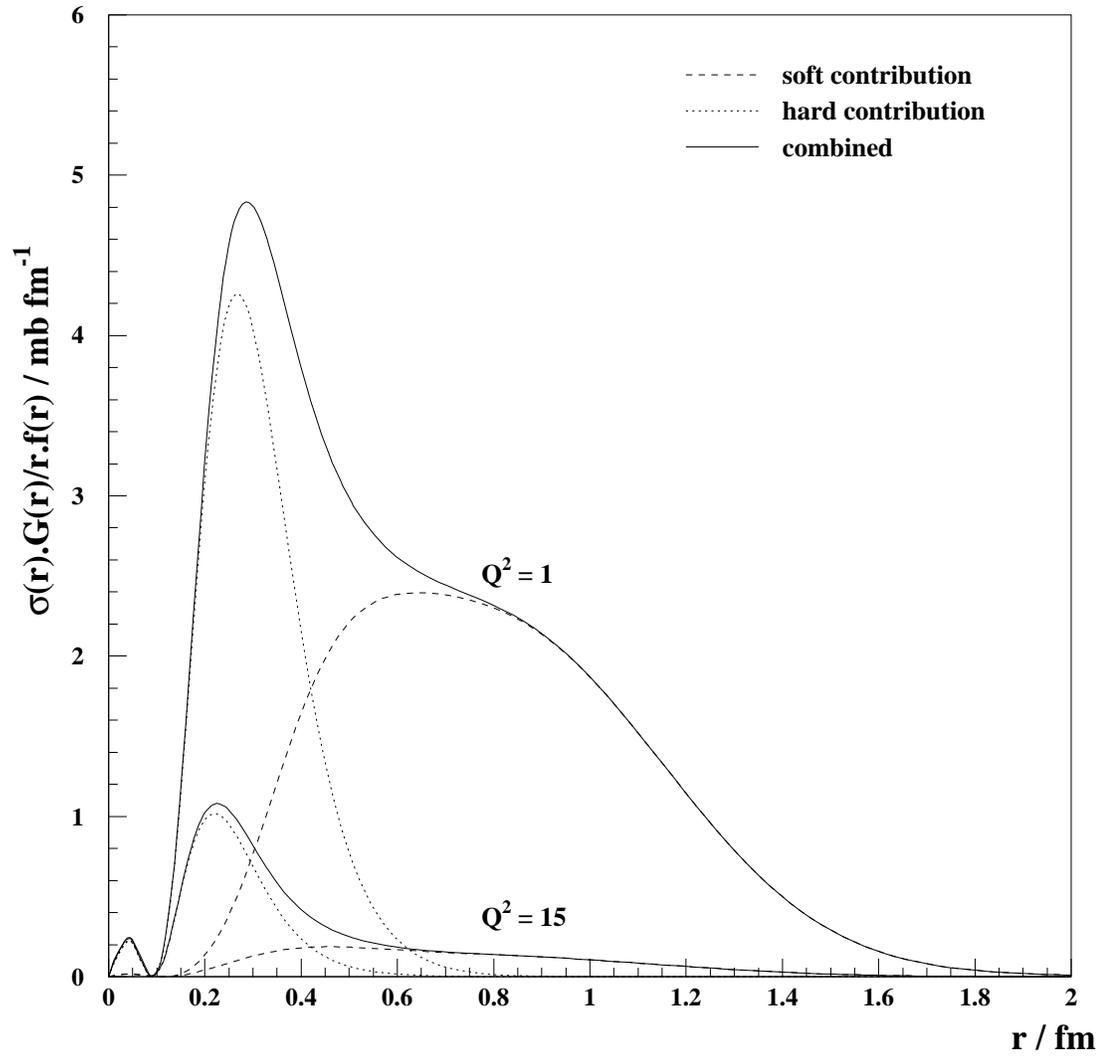}    
    \caption{The relative weighting of the contributions to 
the total photoabsorption cross-sections from dipoles of different
size (fit I).}
    \label{fig:40.1054.3p}
  \end{center}
\end{figure}

\begin{figure}[htbp]
  \begin{center}
   \includegraphics[width=16cm,height=16cm]{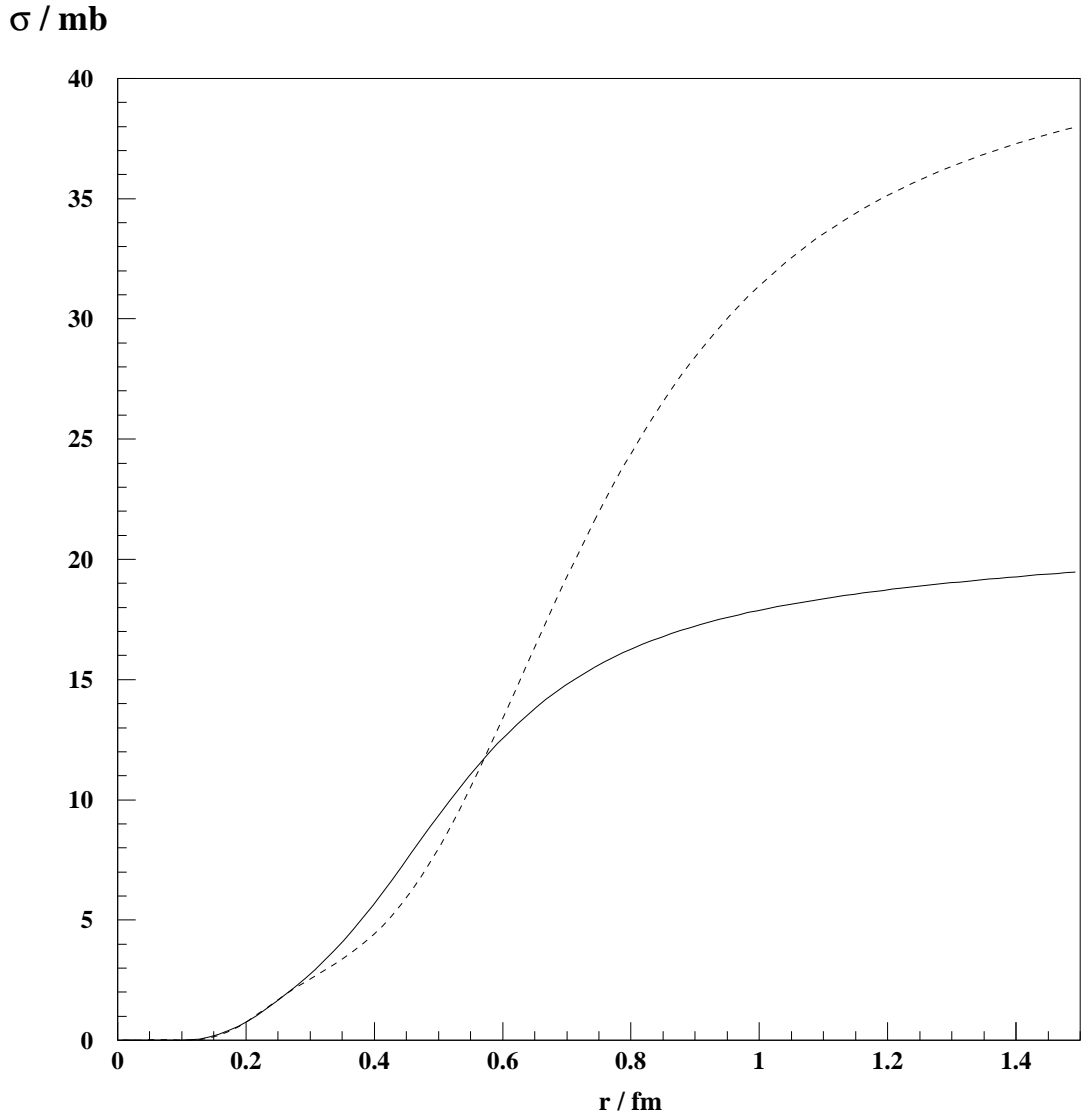}    
    \caption{Comparison of the  dipole cross-section of fit I 
(solid line) with that obtained in 
fit II (dotted  line) at $s = 100$ GeV$^{2}$. The two fits were
constrained to the ZEUS and H1 values for the diffractive ratio
respectively.}
    \label{fig:40.825.657.comp}
  \end{center}
\end{figure}

The energy dependence of the dipole cross-section resulting from Fit I and 
its decomposition into
hard and soft components are shown in Fig.~\ref{fig:40.657.dcs} 
and Fig.~\ref{fig:40.657.hs} respectively.  In addition, 
the contribution to the
total photoabsorption cross-section arising from dipoles of different sizes is
shown in Fig.~\ref{fig:40.1054.3p}, showing that the dipole cross-section is
essentially unconstrained by the data for dipole sizes above about 1.5 fm.
Below this the results accord well with reasonable physical expectations,
with the soft pomeron dominating the large $r$/low $Q^2$ behaviour  and the 
hard pomeron dominating at low $r$/high $Q^2$ when the energy is high enough.
In addition, dipoles of  order 1 fm  have cross-sections commensurate with
typical hadronic cross-sections. The precise value  is 
sensitive to the diffractive ratio imposed in photoproduction, as shown in
Fig.~\ref{fig:40.825.657.comp}, where the dipole cross-sections resulting from 
Fit I (with the ZEUS value imposed) and Fit II (with the H1 value imposed)
are compared.

\subsubsection{The effect of charm}

We have investigated the effect of including a charm  
contribution, assuming dipoles of the same transverse size have the same cross
 section irrespective of flavour. We have adopted a minimalist approach in 
assuming that the
effect of charm flavour on the photon wave function occurs only 
through the charm  mass, leaving the large $r$ peak, $f(r)$, for example, 
unchanged.   
 This leads to an additional term for $G(r)$, of the same form as before but 
with  $m_{f}^{2}$ appropriate to a charmed quark, and weighted $2/3$ in 
accordance with the squared charge coefficient of~(\ref{eq:z_int}). 
The extra term  has the effect 
of increasing the small dipole flux at large $Q^{2}$. 

\begin{figure}[htbp]
  \begin{center}
   \includegraphics[width=16cm,height=16cm]{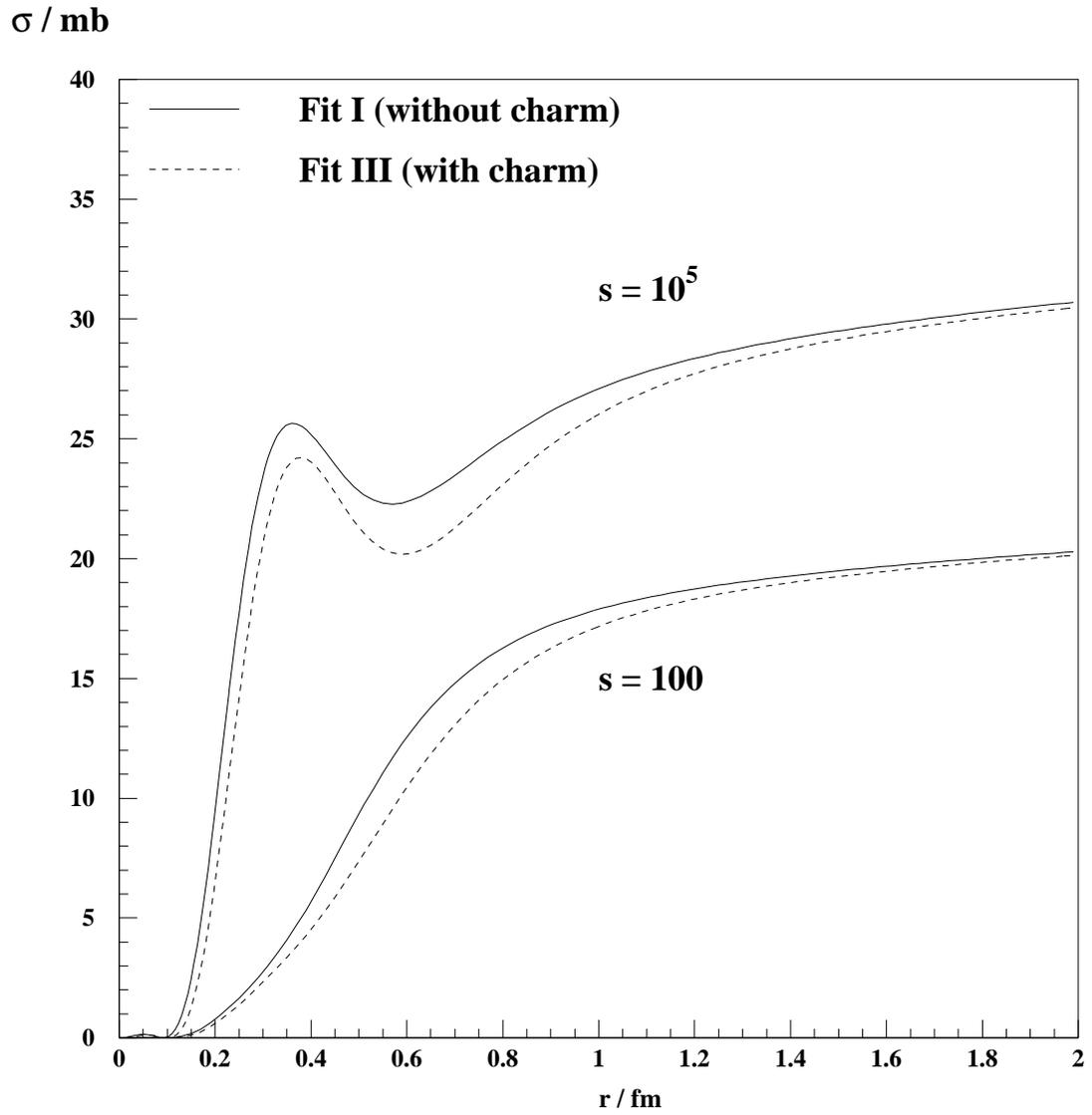}    
    \caption{Comparison at large and small energies of the  dipole 
cross-section of fit I  with that 
obtained in  fit III where a charm contribution was included.}
    \label{fig:41.914.40.1054.dcscomp}
  \end{center}
\end{figure}

Details of a fit 
(fit III) which includes a contribution from charm with an assumed $m_{C}^{2}$
 of 1.4 GeV$^{2}$ are given in Table 3.\footnote{The charm contribution will 
be significant for $Q^{2}$ values in the perturbative region, which makes a 
choice of a running quark mass appropriate. The charm mass is estimated by the
 Particle Data Group to lie
 in the range 1.1 to 1.4 GeV.\cite{cmass_cas}} We have kept the same 
normalisation parameter $a_{0}^{S}$ as fit I to ensure the same diffractive 
ratio. A comparison of the resulting dipole cross section with that of fit I 
is displayed in Fig.~\ref{fig:41.914.40.1054.dcscomp}.  

\begin{figure}[htbp]
  \begin{center}
   \includegraphics[width=16cm,height=16cm]{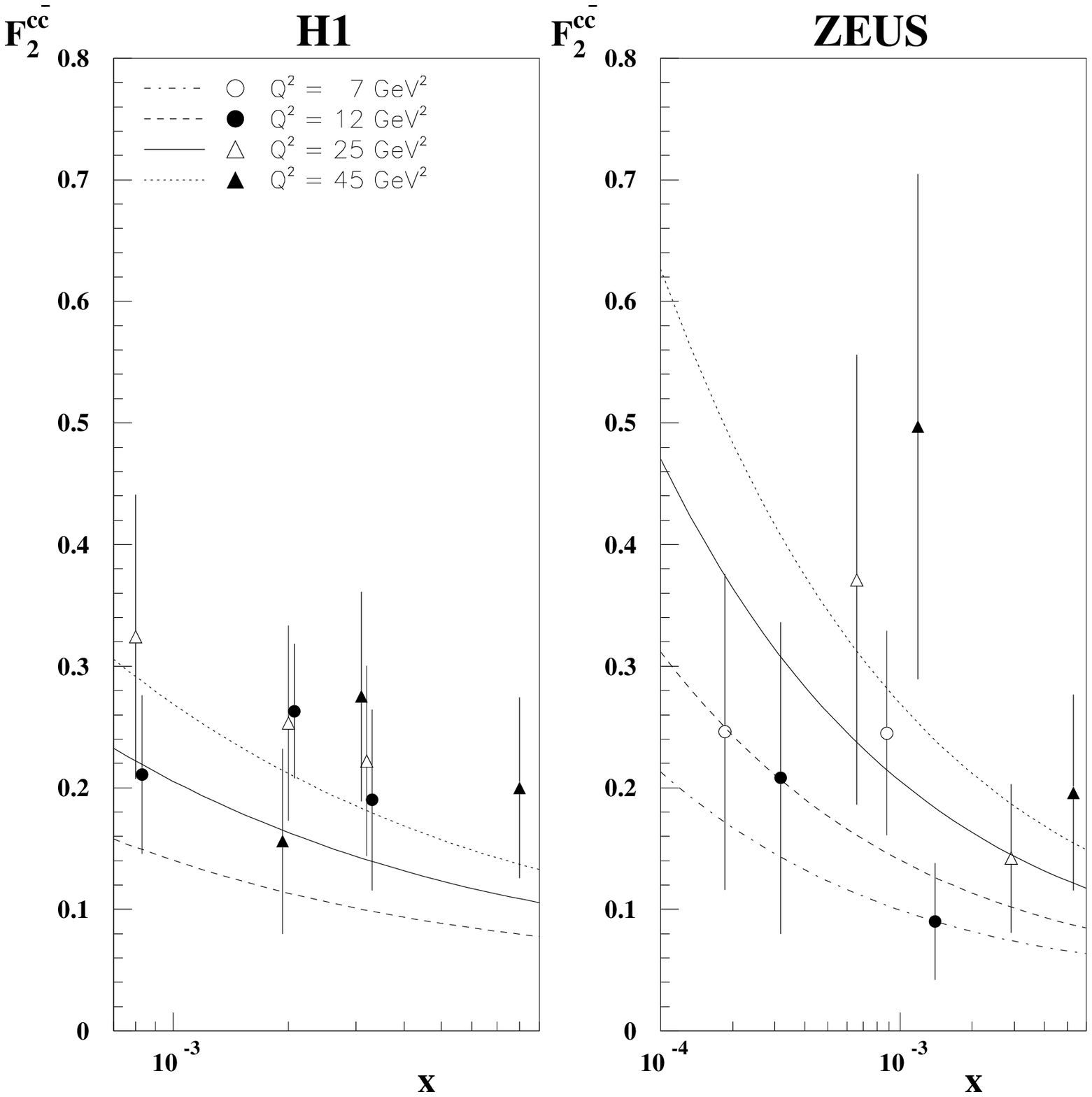}    
    \caption{Comparison of the charm structure function $F_{2}^{c \bar{c}}$ 
predicted from fit III ($m_C^{2} = 1.4 \mbox{ GeV}^{2}$) with experimental 
data.~\cite{dat:H1_charm, dat:ZEUS_charm}}  (Points at the same $x$
have been displaced slightly for clarity.)
    \label{fig:41.914.f2cc}
  \end{center}
\end{figure}

\begin{figure}[htbp]
  \begin{center}
   \includegraphics[width=16cm,height=16cm]{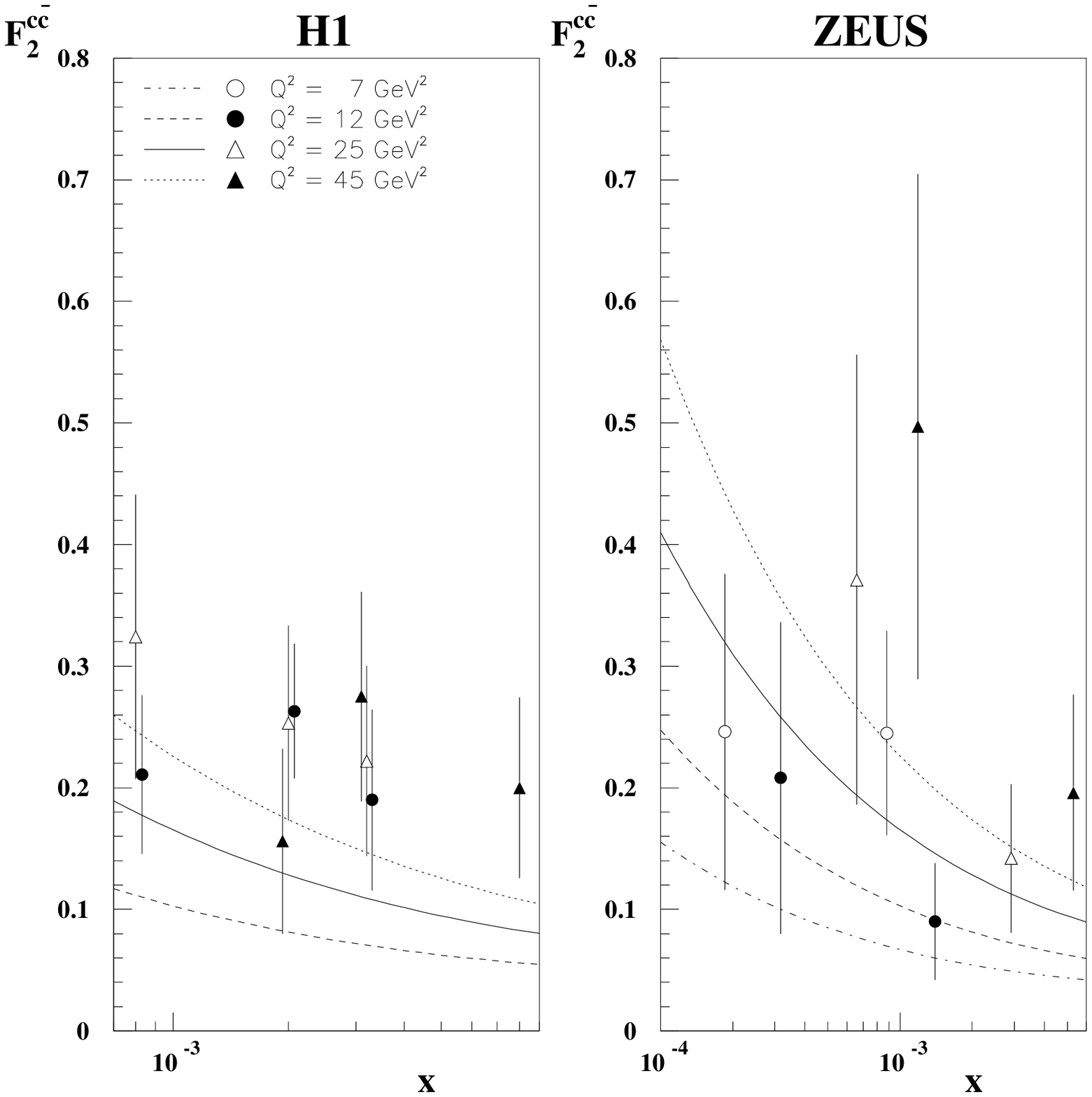}    
    \caption{Comparison of the charm structure function $F_{2}^{c \bar{c}}$ 
predicted from fit IV ($m_C^{2} = 2.3 \mbox{ GeV}^{2}$) with experimental 
data.~\cite{dat:H1_charm, dat:ZEUS_charm}}  (Points at the same $x$
have been displaced slightly for clarity.)
    \label{fig:41.1756.f2cc}
  \end{center}
\end{figure}

As might be expected, the hard term of the dipole cross section is suppressed,
 while the soft term is little affected.  Very little effect on the 
diffractive ratio is observed.   
A comparison of the predicted $F_{2}^{c\bar{c}}$ with data is given in 
Fig.~\ref{fig:41.914.f2cc} showing broad agreement. To gauge the effect of 
increasing the charm mass, we compare predictions from a fit (fit IV) having
 a larger $m_{C}^{2}$ of 2.3  GeV$^{2}$ with the same data in 
Fig.~\ref{fig:41.1756.f2cc}. Within the limitations of the data, the lower 
charm mass value is preferred.

\subsubsection{Other approaches}

A number of other authors have attempted to determine the dipole proton 
cross-section.\cite{sto_vac_pir, vm_dosch_1, dcs_dosch_2, dcs_dosch_1, 
dcs_nik, vm_dcs_nik_1, vm_dcs_nik_2}
The closest in spirit to our approach is that of Golec-Biernat and 
W\"{u}sthoff~\cite{dcs_wust}, who achieve a good fit with a remarkably
simple  parameterisation of the dipole cross-section, which depends on $r$ 
and $x$ rather than $r$ and $s$ as here. Their approach also differs 
from our own in two other  ways. Firstly, they do not fit the  accurate
photoproduction data, so that they are less sensitive to large dipoles
and consequently to confinement effects; and secondly they impose  
saturation at low $x$ (or high $s$) as well as at large 
interquark separations.  Our own success in achieving a fit with 
no saturation in the energy variable indicates that the present data do not 
require it.

\section{Conclusions}
We have succeeded in obtaining a fit to photoabsorption data 
with $Q^2 \le 60$ GeV$^2$, including real photon data,  using a parameterised 
form of the colour dipole cross-section.  This has required modifying the 
effective photon wave function to take account of non-perturbative effects.  
The result is consistent with the cited experimental constraints from
diffractive dissociation  data.

The next step is to develop the model so as to address the more detailed 
diffractive dissociation experimental data in the form of 
the $F_{2}^{D(3)}$ structure function. Also, we should include contributions 
from higher Fock states, such as the $|q\bar{q}g\rangle$, which will dominate 
at low $\beta$~\cite{LC_wust}.

We should also be able to apply our parameterisation to the prediction of 
both vector meson production  and deeply virtual Compton scattering 
cross-sections.

\section{Acknowledgements}

We should like to thank H. G. Dosch and M. W\"{u}sthoff for discussions. 
GK would like to thank PPARC for a Studentship.
This work was supported in part by the EU Fourth Programme `Training and 
Mobility of Researchers', Network `Quantum Chromodynamics and the Deep 
Structure of Elementary Particles', contract FMRX-CT98-0194 (DG 12-MIHT).

\newpage

\section{Tables}
\begin{table}[htbp]
  \begin{center}
Total $\chi^{2}$ 311 (0.88 per d.o.f.)
\[
    \begin{array}{c|c|c|c} 
\hline
 \lambda_{S}  & 0.06 \mbox{ (fixed)} & \lambda_{H} & 0.387 \pm 0.005 \\
            &              &              &    \\ 
 a_{0}^{S}    & 30.05 \mbox{ (fixed)} &              &    \\
             &              &              &    \\
a_{1}^{S}    & 0.12 \pm 0.01   &  a_{1}^{H}   & 0.99  \pm 0.07 \\
             &              &              &    \\
a_{2}^{S}    & -0.202  \pm 0.005  &  a_{2}^{H}   & 0.7 \pm 0.1  \\
             &              &              &    \\
             &              &  a_{3}^{H}   & -6.23 \pm 0.08    \\
             &              &              &    \\
             &              &  \nu_{H}^{2}  & 4.36 \pm  0.02 \\
             &              &              &    \\
B            & 6.4 \pm 0.1  &  c^{2}  & 0.205 \pm 0.004 \\
             &              &              &    \\ 
R            & 6.46 \pm 0.03  &  m^2          & 0.08  \mbox{ (fixed)} \\
\hline  
    \end{array}
\]
Photoabsorption data ($Q^{2} = 0$)
\vspace{0.5cm}
\begin{tabular}{l|c|c}
\hline
Data set & Number of points & $\chi^{2}$ per data point  \\
\hline
Caldwell  & 18 & 1.5  \\
H1        & 1  & 2.1  \\
ZEUS      & 1  & 3.9  \\
\hline
\end{tabular}
    \caption{Fit I, satisfying the ZEUS diffractive ratio for real photons.}
    \label{tab:40.657}
  \end{center}
\end{table}

\begin{table}[htbp]
  \begin{center}
Total $\chi^{2}$ 310 (0.87 per d.o.f.)
\[
    \begin{array}{c|c|c|c} 
\hline
 \lambda_{S}  & 0.06 \mbox{ (fixed)} & \lambda_{H} & 0.380 \pm 0.005 \\
            &              &              &    \\ 
a_{0}^{S}    & 60.28 \mbox{ (fixed)} &              &    \\
             &              &              &    \\
a_{1}^{S}    & 0.032 \pm 0.005   &  a_{1}^{H}   & 0.0  \pm 0.07 \\
             &              &              &    \\
a_{2}^{S}    & -0.094  \pm 0.002  &  a_{2}^{H}   & 7.9 \pm 0.2  \\
             &              &              &    \\
             &              &  a_{3}^{H}   & -13.9 \pm 0.1    \\
             &              &              &    \\
             &              &  \nu_{H}^{2}  & 4.91 \pm  0.02 \\
             &              &              &    \\
B            & 2.40 \pm 0.05  &  c^{2}  & 0.152 \pm 0.005 \\
             &              &              &    \\
R            & 6.08 \pm 0.05  &  m^2          & 0.08  \mbox{ (fixed)} \\
\hline  
    \end{array}
\]
Photoabsorption data  ($Q^{2} = 0$)
\vspace{0.5cm}
\begin{tabular}{l|c|c}
\hline
Data set & Number of points & $\chi^{2}$ per data point  \\
\hline
Caldwell  & 18 & 1.5  \\
H1        & 1  & 1.9  \\
ZEUS      & 1  & 3.8  \\
\hline
\end{tabular}
    \caption{Fit II, satisfying the H1 diffractive ratio for real photons.}
    \label{tab:40.825}
  \end{center}
\end{table}

\begin{table}[htbp]
  \begin{center}
Total $\chi^{2}$ 315 (0.89 per d.o.f.)
\[
    \begin{array}{c|c|c|c} 
\hline
 \lambda_{S}  & 0.06 \mbox{ (fixed)} & \lambda_{H} & 0.380 \pm 0.005 \\
            &              &              &    \\ 
 a_{0}^{S}    & 29.90 \mbox{ (fixed)} &              &    \\
             &              &              &    \\
a_{1}^{S}    & 0.056 \pm 0.008   &  a_{1}^{H}   & 0.47  \pm 0.05 \\
             &              &              &    \\
a_{2}^{S}    & -0.144  \pm 0.004  &  a_{2}^{H}   & 2.5 \pm 0.1  \\
             &              &              &    \\
             &              &  a_{3}^{H}   & -6.56 \pm 0.07    \\
             &              &              &    \\
             &              &  \nu_{H}^{2}  & 4.22 \pm  0.02 \\
             &              &              &    \\
B            & 6.8 \pm 0.1  &  c^{2}  & 0.342 \pm 0.008 \\
            &              &              &    \\ 
R            & 5.67 \pm 0.03  &          &  \\
 m_{L}^2 & 0.08  \mbox{ (fixed)} &  m_{C}^2  
                  & 1.4  \mbox{ (fixed)} \\
\hline  
    \end{array}
\]
Photoabsorption data ($Q^{2} = 0$)
\vspace{0.5cm}
\begin{tabular}{l|c|c}
\hline
Data set & Number of points & $\chi^{2}$ per data point  \\
\hline
Caldwell  & 18 & 1.5  \\
H1        & 1  & 2.2  \\
ZEUS      & 1  & 4.0  \\
\hline
\end{tabular}
    \caption{Fit III, incorporating the charm contribution, with two mass 
squared parameters: $m_{L}^2$ for the light quarks and 
$m_{C}^2$  for the charm quark. The diffractive ratio is the same 
as for fit I.}
    \label{tab:41.872}
  \end{center}
\end{table}

\newpage

\section{Figure Captions}

\begin{trivlist}

\item \textbf{Figure 1} The diffractive process from a mixed position-momentum
 viewpoint. Transverse components are spatial; non-transverse components are 
light cone momenta.
  
\item \textbf{Figure 2}  The weight function $f(r) G(r) / r$ for 
different $Q^{2}$ (fit I). 
The peak at low $Q^{2}$ represents the modification to the photon 
wave function.

\item \textbf{Figure 3}  Representative sample of fitted data points for 
the total cross-section $\sigma_{\gamma p}^{tot}$ compared with curves 
calculated from the parameterised dipole cross-section for different 
 $Q^{2}$ values (fit I).
 
\item \textbf{Figure 4}  Ratio of the overall singly dissociative 
diffractive cross-section to the total cross-section for 
fit I (solid line) and fit II (dotted line).

\item \textbf{Figure 5}  The dipole cross-section at different 
energies (fit I).
 
\item \textbf{Figure 6}  Hard and soft contributions to the dipole 
cross-section (fit I).

\item \textbf{Figure 7}  The relative weighting of the contributions to 
the total photoabsorption cross-sections from dipoles of different
size (fit I).
  
\item \textbf{Figure 8} Comparison of the  dipole cross-section of fit I 
(solid line) with that obtained in 
fit II (dotted  line) at $s = 100$ GeV$^{2}$. The two fits were
constrained to the ZEUS and H1 values for the diffractive ratio
respectively.

\item \textbf{Figure 9} Comparison at large and small energies of the  dipole 
cross-section of fit I  with that 
obtained in  fit III where a charm contribution was included.

\item \textbf{Figure 10} Comparison of the charm structure function 
$F_{2}^{c \bar{c}}$ 
predicted from fit III ($m_C^{2} = 1.4 \mbox{ GeV}^{2}$) with experimental 
data.~\cite{dat:H1_charm, dat:ZEUS_charm}  (Points at the same $x$
have been displaced slightly for clarity.)

\item \textbf{Figure 11} Comparison of the charm structure function 
$F_{2}^{c \bar{c}}$ 
predicted from fit IV ($m_C^{2} = 2.3 \mbox{ GeV}^{2}$) with experimental 
data.~\cite{dat:H1_charm, dat:ZEUS_charm}  (Points at the same $x$
have been displaced slightly for clarity.)

\end{trivlist}

\newpage

\section{Figures}

\end{document}